\documentclass[prd,letterpaper,twocolumn,tightenlines,superscriptaddress,
psfig]{revtex4}

\usepackage{amsmath,amssymb,amsfonts,graphicx,dcolumn}
\usepackage{epstopdf}
\usepackage{color}
\usepackage[utf8]{inputenc}
\usepackage{amsthm}
\usepackage{fontenc}
\usepackage{tikz}
\usepackage[dvips]{hyperref}
\RequirePackage{slashed}
\usepackage{cancel}
\usepackage[normalem]{ulem}

\usepackage{comment}
\usepackage[normalem]{ulem} 
\usepackage{color}
\usepackage{latexsym}
\usepackage{amssymb}
\usepackage{multirow}

\newlength\savedwidth

\begin{document}

\title{GPDs at non-zero skewness in  ADS/QCD model }

\author{Matteo Rinaldi}
\address{Instituto de Fisica Corpuscular (CSIC-Universitat de Valencia), 
Parc Cientific UV, C/ Catedratico Jose Beltran 2, E-46980 Paterna 
(Valencia), Spain.}

\email{mrinaldi@ific.uv.es}

\date{\today}

\begin{abstract}
We study Generalized Parton Distribution 
functions (GPDs) usually measured in hard exclusive processes and encoding 
information on the three dimensional partonic structure of hadrons and 
their spin decomposition, for non zero skewness within the AdS/QCD formalism.
To this aim the canonical scheme to calculate GPDs at zero skewness has been 
properly generalized. Furthermore,
we show that the latter quantities, in this non forward 
regime, are sensitive to non trivial details of the hadronic light front wave 
function, {  such as a kind of parton correlations } usually not accessible 
in  studies of form factors 
and GPDs at zero skewness. 
\end{abstract}

\maketitle

\section{\label{sec:intro}Introduction}

In the last  years, much theoretical and experimental attention has been 
focused on the study of non perturbative quantities in QCD encoding fundamental 
information on the partonic proton structure, e.g., parton distribution 
functions (PDFs), transverse  momentum dependent PDFs (TMDs), double parton 
distribution functions (dPDFs) and generalized parton distribution functions 
(GPDs) \cite{1Pa,5Pa,3Mo,1Mo}. From a 
theoretical point of view, GPDs allow to grasp  information on the three 
dimensional partonic structure of hadrons \cite{2Mo}, and
thanks to the so called Ji's sum rule, to access  the orbital angular momentum 
of partons inside  hadrons. This knowledge is crucial to shed 
some light onto the still open problem of the proton spin crisis. 
From an experimental 
point of view, GPDs represent a challenge due to  the difficulties in the 
measurements of high energy exclusive events (see \cite{6n,7n} 
and references therein).

In this paper, we study the GPDs using the  AdS/QCD  scheme proposed by 
Brodsky and de T\'eramond  \cite{10V1,14Mo,10V2,24Mo,15Mo,17Mo,22Mo} based on 
the  
original AdS/CFT Maldacena conjecture \cite{maldacena}. The idea comes from
 the observation that for small momentum transfer the 
QCD coupling constant  can be approximated by a constant
 and quark masses can been neglected \cite{18Mo}. Confinement can be 
simulated in different ways \cite{10V1,10V2,24Mo,12Mo, katz}. A 
crucial ingredient of this proposal is the mapping between the Light Front (LF) 
Hamiltonian formulation of QCD and the AdS description of hadrons.
This correspondence is implemented
 by relating the fifth dimensional variable $z$ with the LF 
transverse position and longitudinal momentum fractions carried by partons \cite{17Mo}.
This approach has been successfully applied to the calculations of 
the hadronic spectrum, the form factors  and the GPDs for 
$\xi=0$ ($\xi$ represents the skewness) 
\cite{10V1,10V2,24Mo,17Mo,19V,25V,26V,27V,28V,modanff,neww, 
25Mo,gpdsadsqcd,16Mo,19Mo,20Mo,21Mo,23Mo,28Mo,29Mo,30Mo,mondalnew, 26Mo}.
Moreover, a first analysis of hard deep inelastic scattering at small $x$ 
within AdS/CFT has been proposed in Ref. \cite{13Mo}. 
In the present paper we focus our attention in the calculation of the GPDs for 
$\xi\neq0$, within the soft-wall model \cite{10V1,10V2,24Mo},
proposing a suitable extension of the approaches already discussed in 
the literature \cite{24Mo,gpdsadsqcd,trainigpd}.
This step is necessary to provide a complete description of the GPDs in the
AdS/QCD scheme which allows comparison of theoretical results
with data  usually obtained  for $\xi \neq0$. 
Furthermore, our
results open new ways to calculate observables within the AdS/QCD framework
which may lead to useful predictions on the structure of hadrons.

\section{Form Factor in ADS/QCD}
Before we describe GPDs let us review the  AdS/QCD formalism for the 
calculation of form factors
 within the soft-model.  
In this case the breaking of
conformal 
invariance is induced by introducing a quadratic dilaton term 
$\Phi=e^{\pm \kappa^2 z^2}$ in the AdS action. The effect of the breaking can 
be 
directly 
incorporated into the baryonic 
field \cite{18Mo,gpdsadsqcd} and the confining potential, depending on the 
dilaton term, is put in by hand in the soft-wall model.
The strategy to 
 connect AdS quantities to the 
correspondent Light Front (LF) ones is to write
 the AdS Dirac equation,  associated to the AdS action, 
  as a LF  equation and to obtain the corresponding
 baryon wave function by solving an equation in the $2\times2$ spinor 
representation \cite{gpdsadsqcd}. Let us just show the analytic expression of 
such baryonic 
functions  
for two values of the orbital angular momentum, $L_z =0$ and $L_z = 1$, 
respectively \cite{25Mo}:

\begin{eqnarray}
\nonumber
\phi_+(z) &= \dfrac{\sqrt2 \kappa^2}{R^2} z^{7/2}e^{-\kappa^2 z^2/2},
\\
\phi_-(z) &= \dfrac{ \kappa^3}{R^2} z^{9/2}e^{-\kappa^2 z^2/2}
\end {eqnarray}

where $\kappa = 0.4066$ GeV has 
been fixed to fit data on proton and 
neutron form factors \cite{modanff,gpdsadsqcd}.

The other fundamental ingredient of the calculation of  form factors within 
the  AdS/QCD scheme  is the
 bulk-to-boundary propagator whose expression, in the soft wall model, 
reads~\cite{18Mo, 27Mo}:

\begin{align}
\label{gff0}
\nonumber
V(Q^2,z) &= \int_0^1 \tilde V(x,t,z)~ dx  
\\
&=\int_0^1 
   \dfrac{x^{Q^2/(4 
\kappa^2)}}{(1-x)^2} e^{-\kappa^2 
z^2 x/(1-x)}dx  ~.
\end{align}

Following Refs. \cite{18Mo, gpdsadsqcd},
 the final expression for  form factors reads 
as follows:

\begin{align}
\label{f1ads}
 F_\pm(Q^2) = g_\pm R^4 \int \dfrac{dz}{z^4} V(Q^2,z)~ \phi_\pm^2(z)~,
 \end{align}

  where the effective 
charge $g_\pm$ determines the spin-flavor structure of  form factors 
\cite{25Mo,gpdsadsqcd} (see explicit expressions of Dirac and Pauli form 
factors in Ref. \cite{gpdsadsqcd}.)

\section{\label{sec:intro} GPDs in AdS/QCD  }

As already discussed, e.g., in Ref. \cite{gpdsadsqcd}, 
the proton GPDs at $\xi 
=0$, within the soft-wall model, can be calculated from
 form 
factors. However, for the purpose of the present analysis, it is convenient to 
relate the spin dependent light-cone correlator \cite{1Mo,pasquini} 
 $C^q_{\lambda, \lambda'}(x,\xi,t)$  to 
$\phi_\pm(z)$,  being $\lambda~(\lambda')$ the third 
component of the proton spin in the initial (final) state. 
To this aim, using then the parametrization of Dirac and Pauli form factors in 
terms of the flavor dependent GPDs,  one  
obtains \cite{gpdsadsqcd,18Mo}:

\begin{align}
\label{corp}
& C^{u}_{++}(x,0,t) = R^4 \int \dfrac{dz}{z^4} V(x,t,z)\left[\dfrac{5}{3} 
\phi_+^2(z)+\dfrac{1}{3}\phi_-^2(z)  \right]~,
\\
\nonumber
& C^{d}_{++}(x,0,t) = R^4 \int \dfrac{dz}{z^4} V(x,t,z)\left[\dfrac{1}{3} 
\phi_+^2(z)+\dfrac{2}{3}\phi_-^2(z)  \right]
\end{align}

and for the spin flip part one has:

\begin{align}
\label{corm}
\nonumber
& C^{u}_{+-}(x,0,t) = R^4(2\chi_p+\chi_n)  \int \dfrac{dz}{z^4} V(x,t,z)~ 
\phi_-^2(z)~,  
\\
& C^{d}_{+-}(x,0,t) = R^4(2\chi_n+\chi_p)  \int \dfrac{dz}{z^4} V(x,t,z)~ 
\phi_-^2(z) .
\end{align}

Since the light cone correlator directly depends on the LF proton 
wave function \cite{pasquini}, in order to match  results 
found within the AdS/QCD framework with the same 
quantities evaluated within the LF approach, the baryonic functions 
$\phi_\pm(z)$ must be 
related to the LF proton wave functions \cite{17Mo}. To this aim the variable
$z$  turns out a function  of $\vec 
b_{i\perp}$ and  $x_i$, the transverse 
position and longitudinal momentum fraction carried by the 
$i$ parton respectively \cite{17Mo}. 

We proceed to calculate the GPDs for $\xi \neq 0$. These quantities have been 
also studied in a recent paper in Ref. \cite{trainigpd} within a different 
approach and  using the IR improved 
soft-wall 
model \cite{ir}.
To this aim since the LF proton wave function 
is a frame independent quantity \cite{58Br}, it is  useful to work in the 
intrinsic frame where
  $z=z_1 = \sqrt{x_1/(1-x_1)} |\vec b_{1\perp}|$ can be considered as a one 
body variable (for details on the chosen frame see, e.g., Ref. \cite{vento}).
In this scenario,   Eqs. (\ref{corp}, \ref{corm}) can be written in 
terms 
of 
$z_1$, 
associated to the interacting 
parton, and $z_2$ associated to a second spectator particle. 
Within this choice, the light cone correlator can be written:
\begin{align}
\label{cor2}
\nonumber
& C^q_{ab}(x,0,t) = R^8\int dx_1 dx_2 \dfrac{dz_1}{z_1^4}   \dfrac{dz_2}{z_2^4} 
V(x_1,t,z_1)
\\
&\times [n^{q}_{ab} \phi^2_+(z_1,z_2) +  m^q_{ab} 
\phi_-^2(z_1,z_2)]\delta(x-x_1)~, 
\end{align}
 
 being $a,b = \pm$.
As one can see, expressions (\ref{corp},\ref{corm}) are recovered by
choosing $n^{q}_{ab}$ and $m^{q}_{ab}$ according to Eqs. 
(\ref{corp},\ref{corm}) and
properly 
introducing the  following normalization conditions:
\begin{align} \label{1bpwf2}
\nonumber
&  R^4 \int  \dfrac{dz_2}{z_2^4}~ \phi^2_\pm(z_1,z_2) = \phi^2_\pm(z_1);
\\
&  R^8 \int \dfrac{dz_1}{z_1^4} \dfrac{dz_2}{z_2^4}\phi^2_\pm(z_1,z_2) = 1~.
\end{align}

Let us 
call $\phi_\pm(z_1,z_2)$ the
``two body intrinsic'' proton  function, where the spin-flavor part is already 
included.
As will be discussed later on, the expression Eq. (\ref{cor2}) is quite 
suitable for the generalization of the correlator at $\xi \neq0$.
Following the line of Ref. \cite{pasquini},
in order   to include the $\xi$ dependence, it is sufficient to
properly change  
the argument of the two body  intrinsic
proton function, appearing in Eq. (\ref{cor2}). To this aim, let us introduce 
the following variable
$\bar z_i$, in the initial state and  $\tilde z_i$  in the final one:

 \begin{align}
  \label{newz}
  \bar z_i = \sqrt{\dfrac{1-\bar x_i}{\bar x_i} \dfrac{x_i}{1-x_i}  } z_i~,
  ~~
  \tilde z_i = \sqrt{\dfrac{1-\tilde x_i}{\tilde x_i} \dfrac{x_i}{1-x_i}  } 
z_i~;
  \end{align}

where here

\begin{align}
\label{kin1}
 \bar x_1 = \dfrac{x_1+\xi}{1+\xi},~~ \bar x_2 = \dfrac{x_2}{1+\xi}~,
~~
 \tilde x_1= \dfrac{x_1-\xi}{1-\xi},~~ \tilde x_2= 
\dfrac{x_2}{1-\xi}~.
 \end{align}

Starting from the generalization of the correlator,  Eq. 
(\ref{cor2}), in the $\xi \neq 0$ 
case,  using as argument  the variables described in Eqs. 
(\ref{newz},\ref{kin1}), one 
finds:

\begin{align}
 \label{GPDExi}
 \nonumber
& E^{q}_v(x,\xi,t) =  \sqrt{ \dfrac{t_0-t}{-t}} \dfrac{R^8}{\sqrt{ 1-\xi^2}} 
\int dx_1 dx_2 dz_1 dz_2~z_1 z_2
\\
\nonumber
&\times \tilde V(x_1,t,z_1) \dfrac{f_q}{(2\pi)^2} \phi_-(\bar z_1, \bar z_2)
\phi_-(\tilde z_1, \tilde z_2) \dfrac{ \delta(x-x_1) }{[\bar z_1 \bar z_2 
\tilde z_1 \tilde z_2]^{5/2}  
}
\\
&\times\sqrt{\dfrac{\bar x_1 \bar x_2 
\tilde x_1 \tilde x_2  }{(1-\bar x_1)(1-\bar x_2)(1-\tilde x_1)(1-\tilde x_2)
}   } 
 \dfrac{(1-x_1)(1-x_2)}{x_1 x_2}    ~,
 \end{align}

 where, in the last line, functions of $\bar x_i$ and $\tilde x_i$ do not 
cancel the jacobian due to the transformation between $b_{i \perp}$ and $z_i$, 
in this not diagonal case.
Moreover, the flavor coefficient $f_q$ is defined as follows:

 \begin{align}
  &f_u = 2\chi_p + \chi_n~,
~~f_d = 2\chi_n + \chi_p~,
  \end{align}
  
  where $\chi_{p(n)}$ is the anomalous magnetic moment of proton (neutron).
For the GPD $H$ one finds:

\begin{align}
  \label{GPDHxi}
& H^{q}_v(x,\xi,t) =  \dfrac{R^8}{{ 1-\xi^2}} 
\int dx_1 dx_2 dz_1 dz_2~z_1 z_2 
\\
\nonumber
&\times \tilde V(x_1,t,z_1)
 \left[    
\dfrac{\xi^2}{\sqrt{1-\xi^2}}   \sqrt{\dfrac{t_0-t}{-t}} \phi_-(\bar z_1, \bar 
z_2)
\phi_-(\tilde z_1, \tilde z_2)f_q +\right.
\\
\nonumber
&+\left.  F_q(\bar z_1, \bar z_2,\tilde z_1, \tilde z_2)    \right. \Bigg] 
\dfrac{1}{(2\pi)^2} \dfrac{1}{[\bar z_1 \bar z_2 \tilde z_1 \tilde 
z_2]^{5/2}  
}\delta(x-x_1)
\\
\nonumber
&\times
\dfrac{(1-x_1)(1-x_2)}{x_1 x_2}
\sqrt{\dfrac{\bar x_1 \bar x_2 
\tilde x_1 \tilde x_2  }{(1-\bar x_1)(1-\bar x_2)(1-\tilde x_1)(1-\tilde x_2)
}   } 
~.
\end{align}

where in this case the following flavor function has been introduced:

\begin{align}
\nonumber
 F_q(z_1,z_2,z_2,z_4) &= 
\left[n^q_{++}\phi_+^2(z_1,z_2)+m^q_{++}\phi_-^2(z_1,z_2)         
\right]^{1/2}
\\
&\times \left[n^q_{++}\phi_+^2(z_3,z_4)+m^q_{++}\phi_-^2(z_3,z_4)       
 \right]^{1/2}~.
\end{align}

\section{Numerical solutions for the two body intrinsic function}

In order to evaluate GPDs at $\xi\neq0$ the calculation of the two body 
intrinsic function is necessary, 
however,  the only constraint for  the evaluation of 
$\phi_\pm(z_1,z_2)$  is the  integral 
Eq. 
(\ref{1bpwf2}). Due to this condition different two body intrinsic 
distributions, corresponding to    
the same 
1 
body 
one ($\phi_\pm(z)$), can be  found. In particular, in the present 
analysis, two physically different 
scenarios have been considered, i.e. a fully uncorrelated ansatz, $ 
_1\phi_\pm(z_1,z_2)$ and a correlated one, $ 
_2\phi_\pm(z_1,z_2)$.
For the  uncorrelated case, one can 
straightforwardly  
consider a function where the $z_1$ and $z_2$ dependence is fully factorized:

\begin{eqnarray} \label{pwf1p}
 _1\phi_\pm(z_1,z_2) = \phi_\pm(z_1)\phi_\pm(z_2)~.
\end{eqnarray}

For the evaluation of the correlated plus distribution
$_2\phi_+(z_1,z_2)$,  a
numerical solution  of 
Eq. (\ref{1bpwf2}) has been used:
\begin{align}\label{pwf2p}
& _2\phi_+(z_1,z_2) = \dfrac{C_5}{R^4}     e^{-\kappa^2 \beta(z_1,z_2)/2} 
\kappa^{5} z_1^4 z_2^4  
\\
\nonumber
& \times
\left. \Big[ 
 C_1 U^2(-1,1,C_4~ \beta(z_1,z_2))+ C_2 U^2(-2,1,C_4~ \beta(z_1,z_2)) \right. 
\\
\nonumber
&+\left.  C_3 U(-1,1,C_4~ \beta(z_1,z_2)) U(-3,1,C_4~ \beta(z_1,z_2)) 
\right.\Big]^{1/2} 
\end{align}

being~ $\beta(z_1,z_2) = z_1^2 + z_2^2 + z_1 z_2 $ and $U(a,b,z)$  the Tricomi 
confluent hypergeometric function  and the coefficients read:

\begin{align}
\nonumber
&C_1 = 1.68944~;
~~C_2 = 17.977~;
~~C_3 = -18.5158~;
\\
&C_4 = 0.082599~;
~~C_5 = 2.75434~.
\end{align}

\begin{figure*}[t]
\includegraphics[scale=0.4]{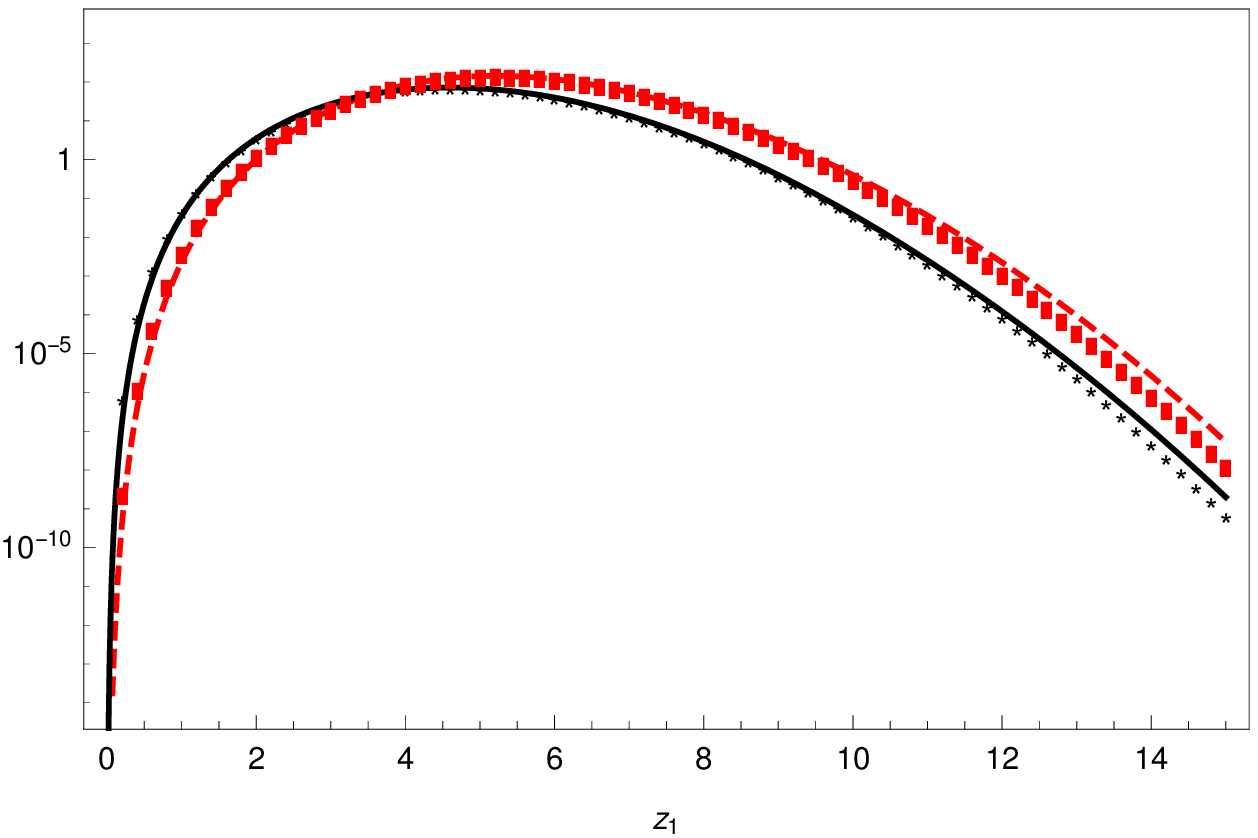}
\hskip 0.5cm
\includegraphics[scale=0.4]{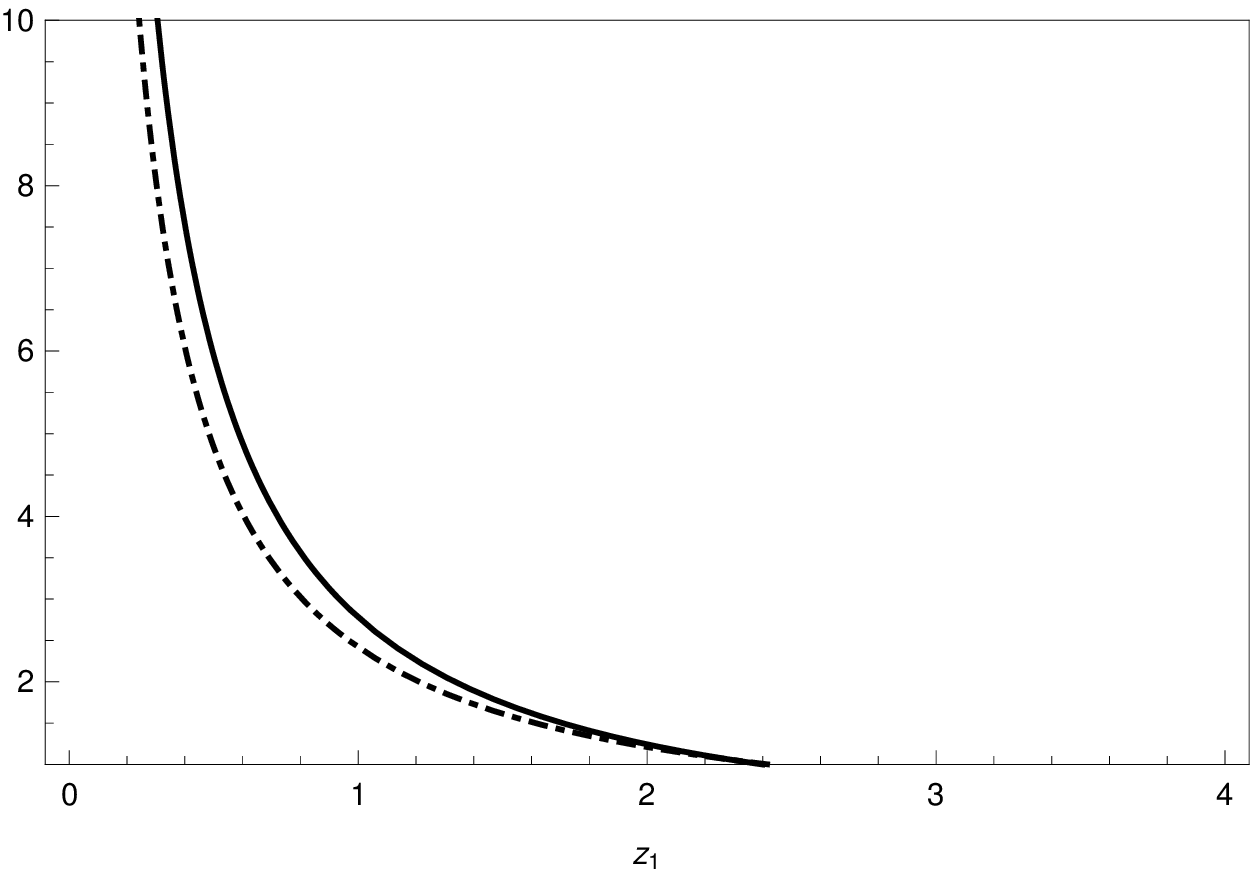}
\hskip 0.5cm \includegraphics[scale=0.4]{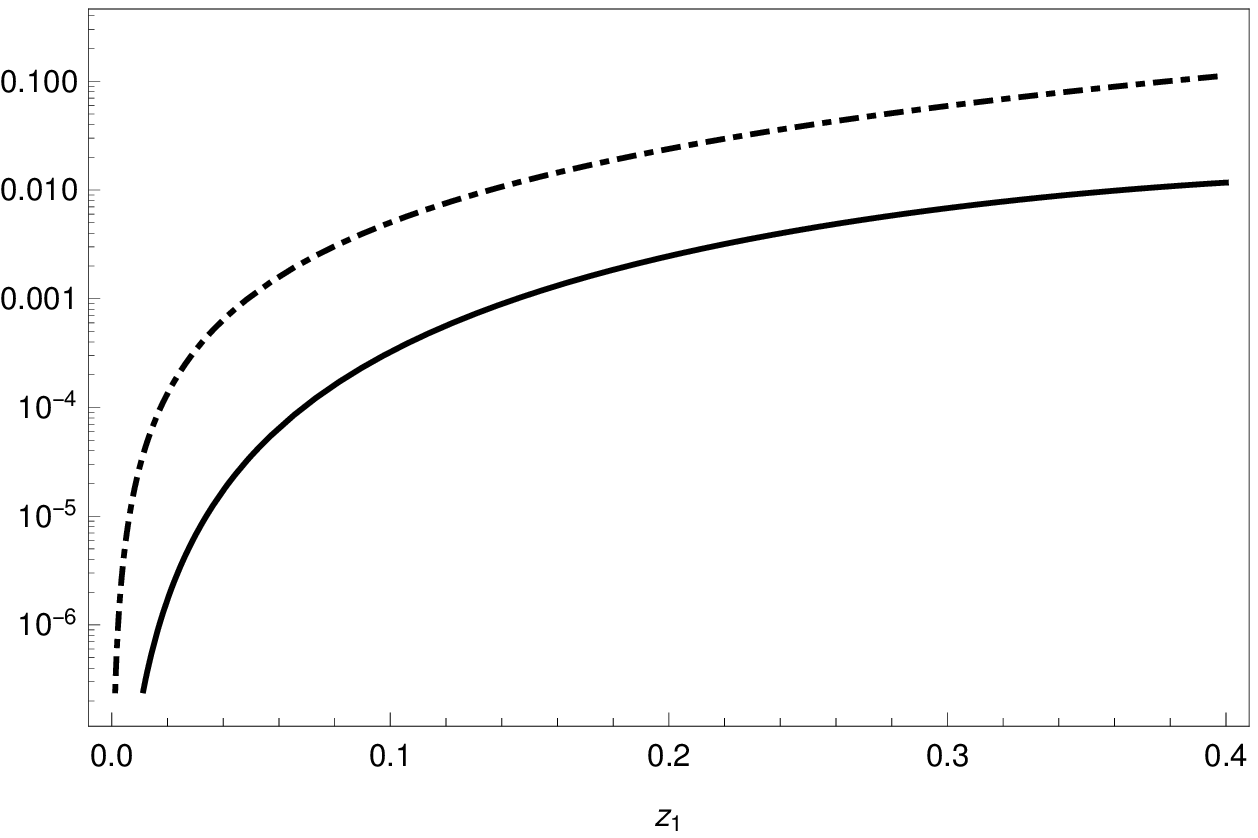}
\caption{\footnotesize  \textsl{Left panel:
Comparison between the
analytic functions $\phi^2_\pm(z_1)$ and the numerical integral
 Eq. (\ref{1bpwf2}) evaluated by means of the correlated two body intrinsic 
function. 
Full line integral over $z_2$ of $_2 \phi_+(z_1,z_2)$. Star points 
$\phi_+(z_1)$. Dashed line integral over $z_2$ of $_2 \phi_-(z_1,z_2)$. 
Squared points 
$\phi_-(z_1)$.
 Central panel: Full line the ratio 
$_2\phi_+(z_1,z_2=5~\mbox{GeV}^{-1})/_2\phi_-(z_1,z_2=5~\mbox{GeV}^{-1})$, 
dot-dashed line the ratio 
$_1\phi_+(z_1,z_2=5~\mbox{GeV}^{-1})/_1\phi_-(z_1,z_2=5~\mbox{GeV}^{-1})$.
Right panel:
the difference between the calculations of the two body intrinsic functions 
within the  
correlated and uncorrelated ansatz. Full line: 
$_1\phi_+(z_1,z_2=5~\mbox{GeV}^{-1} )-_2\phi_+(z_1,z_2=5~\mbox{GeV}^{-1} )$;
dot-dashed line: $_1\phi_-(z_1,z_2=5~\mbox{GeV}^{-1} 
)-_2\phi_-(z_1,z_2=5~\mbox{GeV}^{-1} )$.  }}
\label{intpwf}
\end{figure*}

\noindent
For the minus  component, one finds
\begin{align}
 &_2\phi_-(z_1,z_2) = \dfrac{D_4}{R^4}   e^{-\kappa^2 
\beta(z_1,z_2)/2}  
\kappa^{7} z_1^{5} z_2^{5}
\\
\nonumber
&\times \left.\Big[ D_2 U^2(-1,1,D_1~ \beta(z_1,z_2))+ 
 D_3 U^2(-2,1,D_1~ \beta(z_1,z_2)) \right. 
\\
\nonumber
&+ \left. ~ D_5 U(-1,1,D_1~ \beta(z_1,z_2)) 
 U(-3,1,D_1~ \beta(z_1,z_2)) \right.\Big]^{1/2} 
\end{align}

where now the coefficients read:

\begin{align}
\nonumber
& D_1 = 1.48814~;
 ~~D_2 = 38.6679~;
~~D_3 = 0.208628~;
\\
&D_4 = 0.04298~;
 ~~D_5 = 3.73552~.
\end{align}

In order to qualitatively show the accuracy of the procedure, in  Fig. 
\ref{intpwf},  the integrals over $z_2$ of 
$_2\phi^2_+(z_1,z_2)$  ( full line) and $_2\phi^2_-(z_1,z_2)$ (dashed
line) 
are compared to $\phi^2_+(z_1)$ ( star points) and $\phi^2_-(z_1)$ (squared
points) respectively. As one can see the 
accuracy is quite good for $z_1<10$ GeV$^{-1}$.
With the comfort of these successful checks, in the next section,  
 the result of the 
calculation of the GPDs at $\xi \neq 0$, within the correlated and uncorrelated
ansatz, will be discussed.

\begin{figure*}[t]
\begin{center}
\includegraphics[scale=0.45]{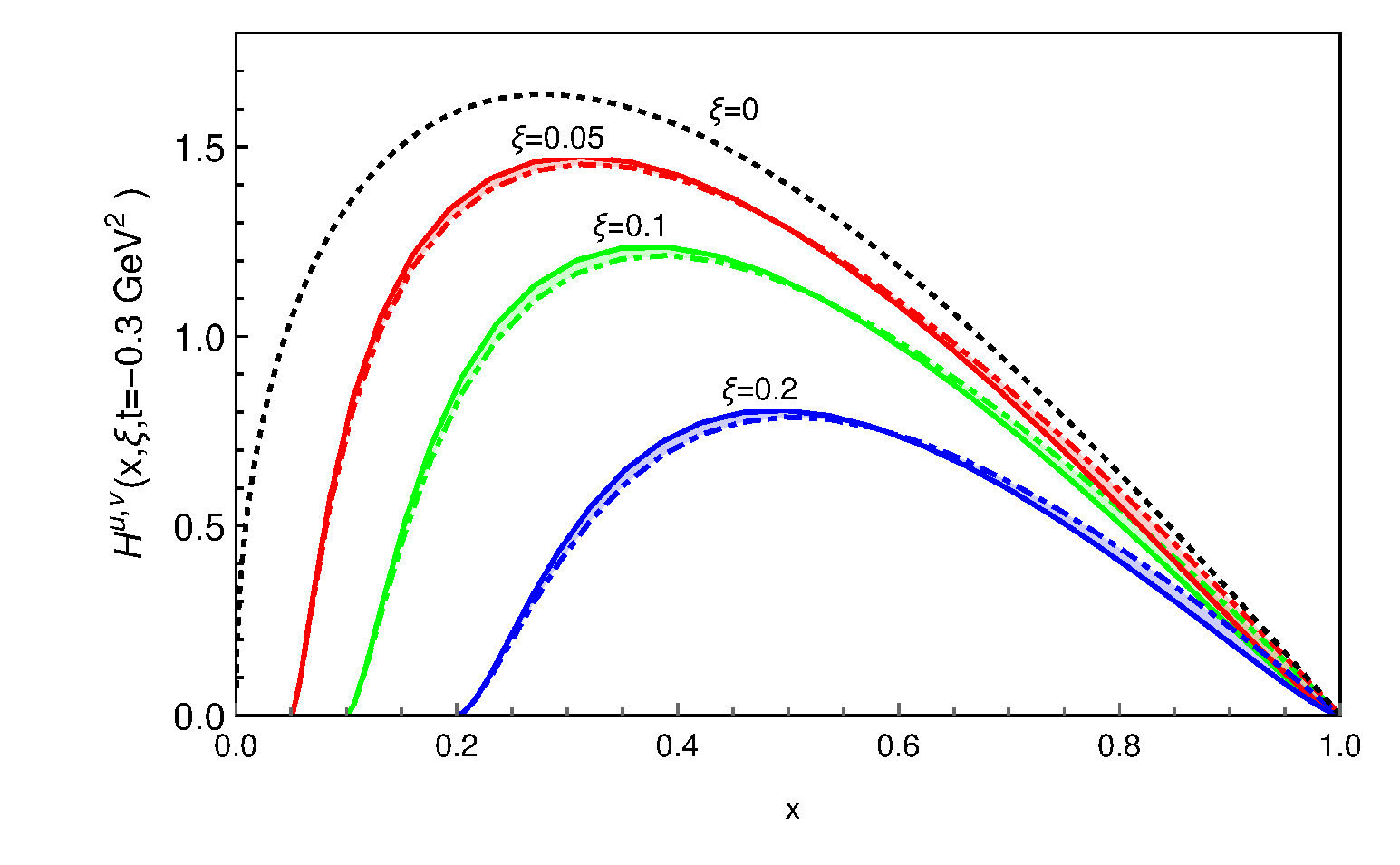} 
\hskip 1.5cm\includegraphics[scale=0.53]{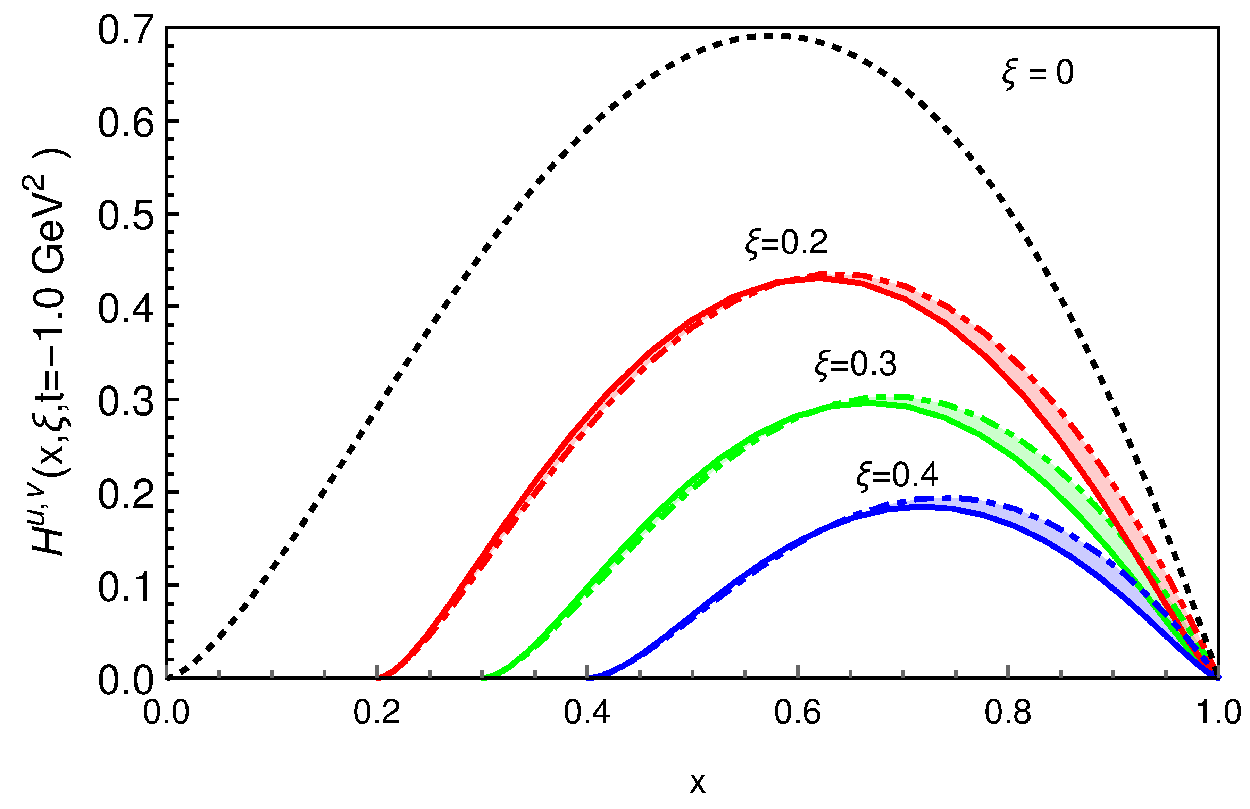} 
\caption{ \footnotesize  \textsl{Left panel: the GPDs $H^u_v(x,\xi,t)$ 
evaluated at 
$t=-0.3$ 
GeV$^2$ for four values of
$\xi=0, 0.05, 0.1,0.2$. The full line is 
obtained by means of the correlated two body intrinsic function,  the 
dot-dashed 
lines is obtained by means of the 
uncorrelated one and dotted is obtained for $\xi=0$ (same result obtained and 
discussed in Ref. \cite{gpdsadsqcd}). The band between the two lines represent 
the theoretical 
error.
Right panel: same quantity of left panel evaluated at 
$t=-1$ 
GeV$^2$ and
$\xi=0,0.2,0.3,0.4$.}}
\label{H03}
\vskip 0.5cm
\includegraphics[scale=0.55]{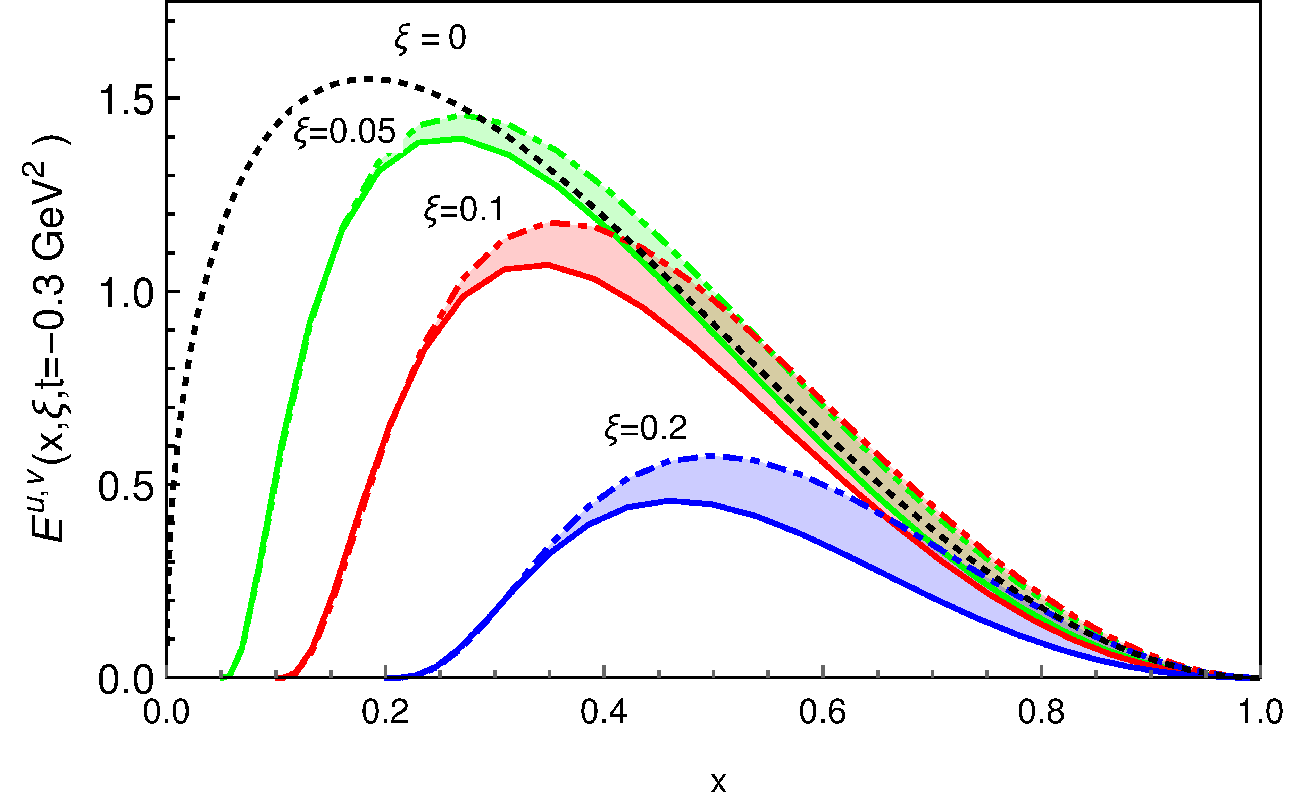} 
\hskip 1.5cm\includegraphics[scale=0.56]{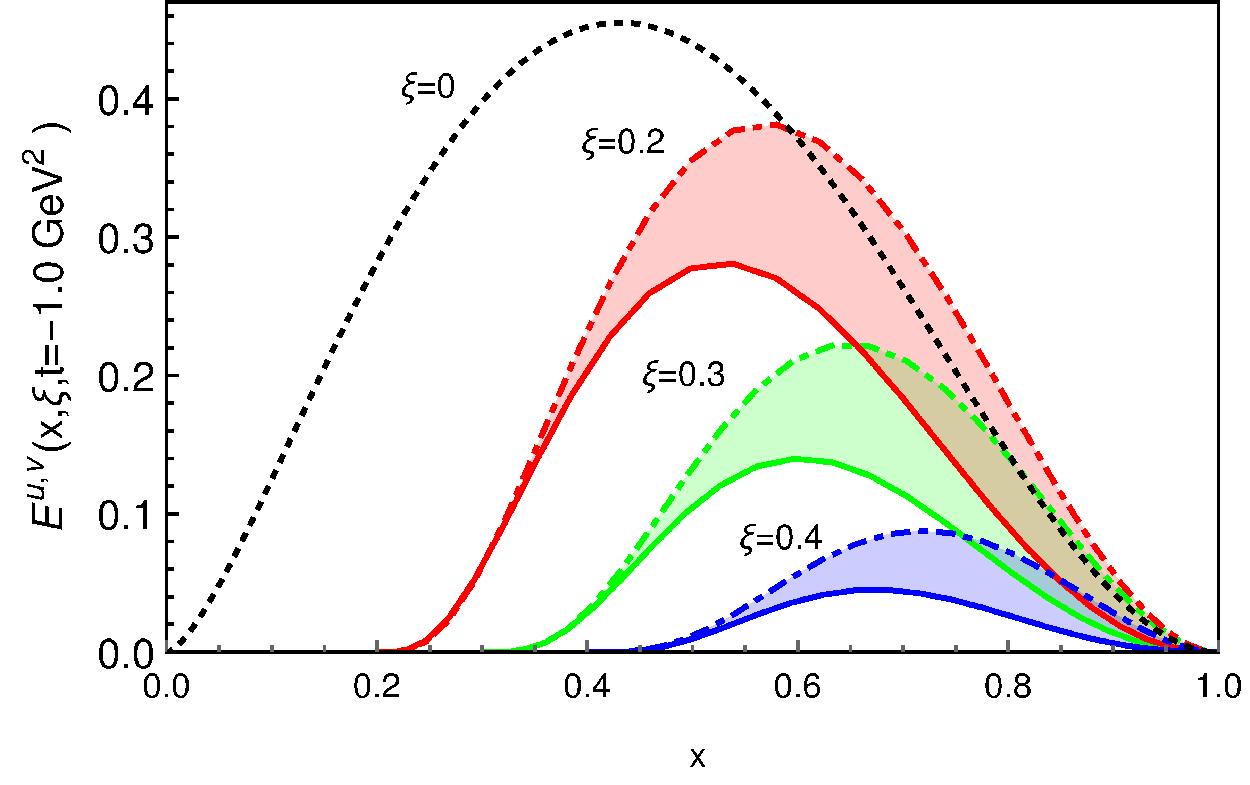} 
\caption{ \footnotesize  \textsl{ Same of Fig. \ref{H03} but for
the GPDs $E^u_v(x,\xi,t)$.}}
\label{E03}
\end{center}
\end{figure*}

\section{Numerical analysis of the GPDs at $\xi \neq 0$}

In this section, the main results of the calculations of the valence GPDs at 
$\xi 
\neq 0$ are presented. The difference between the calculations of the GPDs,  
performed by means of the uncorrelated and 
correlated scenarios can be considered as the theoretical error of the 
present approach.
In  Fig. 
\ref{H03},  the GPD $H^u_v(x,\xi,t)$ has been 
shown for $t = -0.3$ GeV$^2$  (left panel) and $t = -1.0$ GeV$^2$ (right panel)
for four values of $\xi$ (see caption of  Fig. 
\ref{H03}). 
As one can notice, the shape of GPDs, in the $\xi$ dependence, 
is basically decreasing, qualitatively in agreement with results discussed in 
Refs. 
\cite{pasquini, vento}, where GPDs have been calculated within 
constituent quark 
models.
This trend is also confirmed for the GPD $E^{u}_v(x,\xi,t)$, see 
Fig. 
\ref{E03}.
Same results are found for the flavor $d$.
Moreover, for $\xi =0$, as expected, results discussed in Ref. 
\cite{gpdsadsqcd} 
are fully recovered using both the correlated and uncorrelated form of 
$\phi_\pm(z_1,z_2)$. Furthermore, for the GPDs $H$, one 
can notice that
the difference between the calculations with the correlated 
and the uncorrelated two body intrinsic proton  functions is quite small. This 
feature 
can be understood by looking at the central panel of  Fig. \ref{intpwf} where 
the ratio  
$_{1(2)}\phi_+(z_1,z_2)/_{1(2)}\phi_-(z_1,z_2)$ has been shown for $z_2=5$ 
GeV$^{-1}$ (approximately the maximum of the distributions). The full line 
corresponds to the calculation with the 
correlated ansatz and the dot-dashed one is obtained by means of the 
uncorrelated two body intrinsic function. As one can see, for small values of 
$z_1$ in 
both cases the plus component is dominant w.r.t. the minus one. Let us remark 
that since 
 the 
bulk-to-boundary propagator, Eq. (\ref{gff0}), is peaked around $z_1 \sim 0$,
 the integrals Eqs. (\ref{GPDExi}, \ref{GPDHxi}) are   dominated by the small 
$z_1$ region. Furthermore, in this range of $z_1$, the difference between the 
correlated and 
uncorrelated calculations of the plus component of the two body intrinsic 
function (see 
the full line in the right panel of 
Fig. \ref{intpwf})
  is dramatically small, a feature that explains the 
small error band in the evaluation of the GPD $H$ in  
Fig. 
\ref{H03}.
Regarding the GPD $E$, since in this case, as one can see in Eq. 
(\ref{GPDExi}), this quantity depends only on the minus component of the two 
body intrinsic proton function,  the error between the correlated and 
uncorrelated ansatz is bigger then in the plus component case.
One can realize such feature by looking at
the dot-dashed line in the right panel of Fig. \ref{intpwf}. This aspect 
explains why, in the same kinematic conditions, the error in the calculation of 
the GPD $E$ is bigger then that in the  GPD $H$ case.
Let us stress that for both GPDs for very small values of 
$\xi$ and $t$, the kinematic condition useful to study  Ji's sum rule and 
the  three dimensional partonic structure of the 
proton, the error in the present 
approach is quite small. 
In particular, in this framework, the  decreasing trend of the GPDs in the 
$\xi$ dependence is due to partonic correlations.
Furthermore, our analysis shows that if
  high values 
of $\xi$ and $t$ are reached in  experiments, the GPDs at $\xi\neq0$ 
are more sensitive to  details of 
the full proton wave function than form factors or GPDs at $\xi=0$, allowing to 
access  parton 
correlations usually integrated out in the diagonal case.
{  This analysis has taught us how distributions sensitive to correlations, 
like GPDs in non forward regions, evaluated thanks to the experience gained in 
the LF and AdS/QCD approaches,  can be used to find 
the 
importance of correlations in the structure of hadrons. In further studies, 
this analysis will be completed by 
calculating, within the present scheme,
other distributions, like dPDFs already investigated
within the LF approach 
\cite{jhp1,jhp2,plb}, and which have shown to be sensitive to the kind of
correlations here addressed. }
 Work is in 
progress in that direction. A first evaluation of  an approximated 
expression
of dPDFs, through the AdS/QCD correspondence,  together  with the calculation 
of an experimental 
observable
is discussed in Ref. \cite{effA}.

\section{Conclusion}
In the present analysis, GPDs  have been calculated in a fully non 
forward region, namely $\xi$ and $t$ different from zero. To this aim,  the 
usual strategy, developed to evaluate GPDs at $\xi=0$, within the AdS/QCD 
correspondence together with the soft-wall model, has been extended in 
order to evaluate the full proton Light-Front wave 
function from AdS/QCD, including, in principle, 
 two body correlations. As 
shown, within this approach, results previously discussed in other analyses 
have been successfully recovered and the  $\xi$ dependence found for the leading 
twist GPDs $H$ and $E$, evaluated for different flavors, is compatible with the 
one discussed in calculations with constituent quark models. 
Furthermore, since in the present study the full proton wave function is 
obtained by solving an integral equation, different solutions have been 
scrutinized for the numerical evaluations of GPDs and a discussion 
on the theoretical error of the approach has been provided. In particular, 
for small values $\xi$, using different proton wave functions,
leading to same form factors and GPDs at $\xi=0$, similar results have 
been found. However, since at high values of $\xi$, in particular for the GPD 
$E$, these quantities start to be sensitive to details of the 
 proton structure, e.g.,  to two body correlations,   differences in the 
calculations of the latter with correlated and uncorrelated distributions 
become sizable.
Results presently discussed demonstrate that in principle new information on 
 partonic structure of hadrons can be obtained from GPDs at $\xi \neq0$, 
indirectly accessing  two parton correlations, usually studied, e.g. with 
double parton distribution function in double parton scattering. In closing,
our analysis shows that AdS/QCD can be used 
in the future 
to estimate other fundamental observables and parton distributions.

\section*{Acknowledgements}
This work was supported  by  Mineco under contract 
FPA2013-47443-C2-1-P and SEV-2014-0398.
We warmly thank Sergio Scopetta, Vicente Vento and Marco Traini for 
many useful discussions.

\end{document}